# Influence on superconductivity in the parity mixing superconductor Li$_2$T$_3$B(T:Pt,Pd) by non-magnetic impurity and defect doping

G. Bao$^{a*}$, Y. Inada$^{a,b}$, G. Eguchi$^c$, Y. Maeno$^c$, M. Ichioka$^a$, G-q Zheng$^a$

$^a$Department of Physics, Okayama University, Okayama 700-8530, Japan
$^b$Faculty of Education, Okayama University. Okayama 700-8530, Japan
$^c$Department of Physics, Graduate School of Science, Kyoto University. Kyoto 606-8501, Japan

Superconductors with noncentrosymmetric crystal structures such as Li$_2$T$_3$B(T:Pd,Pt) have been the focus of in-depth research with their parity mixing nature. In this study, we focused our research on non-magnetic impurity effect in Li$_2$T$_3$B (T: Pd, Pt). The nature of the pair breaking by non-magnetic impurity in the parity mixing superconducting state is still unclear. We prepared different quality samples of Li$_2$Pd$_3$B and Li$_2$Pt$_3$B by changing conditions in synthesizing, and sample qualities were estimated by residual resistivity. Spin singlet dominant superconductor Li$_2$Pd$_3$B exhibits the weak $T_c$ suppression attributed by nonmagnetic impurity and defects, while H$_{c2}$ (0) value increased. This behavior is similar in ordinary **s**-wave superconductor. On the other hand, for the spin triplet dominant superconductor Li$_2$Pt$_3$B, it was suggested that the Cooper pair was broken and superconducting gap was decreased by non-magnetic impurities and defects. Li$_2$Pt$_3$B is similar to unconventional superconducting state.

## 1. Introduction

The superconductors with noncentrosymmetric crystal structures are the focus of attentions. In superconductors with ordinary crystal structures with an inversion center, superconducting paring function has definite parity, spin singlet for odd parity and spin triplet for even parity. On the other hand, the broken inversion symmetry in the crystal causes finite antisymmetric spin orbit coupling (ASOC) [1-3], which lead to unique superconducting properties. Superconductivity cannot be classified in terms of the spin-singlet or the spin-triplet state in the system. It is also argued that the superconducting order parameter can be formed as a mixture of these two components, even without a spin-triplet term in the pairing interaction [4,5].

The first discovery of noncentrosymmetric heavy fermion unconventional superconductor CePt$_3$Si opened the new physics in superconductivity [6]. Many superconductors with noncentyosymmetric crystal structures have been identified among heavy fermion systemes such as UIr [7], CeRhSi$_3$ [8,9], CeIrSi$_3$ [10,11], CeCoGe$_3$ [12], and others like Li$_2$Pd$_3$B, Li$_2$Pt$_3$B, and Li$_2$(Pd$_{1-x}$Pt$_x$)$_3$B [14,23, 27].

Li$_2$Pd$_3$B and Li$_2$Pt$_3$B crystalize with same noncentrosymmetric crystal structures (P4$_3$32 in space group) [13]. They have attracted particular attention, because they exhibit completely different superconducting properties [14-18]. Li$_2$Pd$_3$B behaves as a full-gap superconductor, while Li$_2$Pt$_3$B has line nodes in the energy gap [15-20]. Previous studies of penetration depth [17], NMR [15,16], and specific heat measurements [18] suggested that Li$_2$Pt$_3$B is a spin triplet dominant superconductor. The ratio of singlet-like to triplet-like order parameter was estimated as 0.6, with the inclusion of sufficiently large amount of spin-singlet-like order parameter components, while Li$_2$Pd$_3$B is an s-wave spin singlet dominant superconductor with the ratio of 4 [17]. We expect they are candidate to study the parity mixing superconductors. Furthermore, no strong electron correlation is observed in Li$_2$Pd$_3$B and Li$_2$Pt$_3$B, as contrasted with CePt$_3$Si. CePt$_3$Si has a tetragonal crystal structure with no inversion center due to the lack of the mirror plane perpendicular to the only one axis (c-axis), while Li$_2$Pt$_3$B is in a cubic structure composed of distorted octahedral units of BPd$_6$ or BPt$_6$. There is no mirror plane in all directions and no inversion center. These differences may be important in the discussion of superconductivity in noncentrosymmetric crystal structures with ASOC without strong electron correlation.

In the NMR experimental study of Li$_2$(Pd$_{1-x}$Pt$_x$)$_3$B, it was reported that the paring symmetry changes drastically at $x$=0.8 [20]. For x≤0.8, the materials are in a predominantly spin-singlet state, while the x>0.8, unconventional properties due to the mixing of the spin triplet state appear. The change is caused by an abrupt enhancement of the ASOC due to an increased distortion of the B (Pd, Pt)$_6$ octahedral units [20].

It is known that s-wave superconductor is not affected by non-magnetic impurity and defects doping in contrast to that of a non s-wave superconductor. It is caused by the sign inversion of the order parameter on the Fermi surface, although the nature of the pair breaking by non-magnetic impurity in the parity mixing superconducting state is still unclear. We expect that parity mixing ratio can be controlled by non-magnetic impurity effect. We investigated the non-magnetic impurity effect in $Li_2T_3B$ (T: Pd, Pt). In this paper, we present that the non-magnetic impurity effect in $Li_2Pd_3B$ and $Li_2Pt_3B$, by changing conditions in synthesizing. We investigated the relations of residual resistivity and $T_c$ as well as their superconducting phase diagram.

Impurity effects in noncentrosymmetric crystal structures are discussed in theoretical calculations [21, 22]. The two-component structure of the order parameter allows for two distinct pairing channels, a dominant and a subdominant gap. In the dirty system only the conventional pairing component would eventually survive, while all alternative pairing channels are suppressed.

In this paper, we present the non-magnetic impurity effect in $Li_2Pd_3B$ and $Li_2Pt_3B$.

## 2. Experimental

Polycrystalline samples were prepared by the two-step arc melting method [23]. In the first step, $Pd_3B$ or $Pt_3B$ was synthesized by using Pd (99.95%), Pt (99.999 or 99.99%), B (99.5%). In the second step, excess Li (99.9%) of 10% has been introduced into the $Pd_3B$ or $Pt_3B$ alloys. The melting points of Pd, Pt and B are higher than the boiling point of Li. It causes grave difficulty in synthesizing these compounds. $Pd_3B$ and $Pt_3B$ have lower melting points than the boiling point of Li, so the two-step arc melting method is effective to avoid this difficulty. We prepared different quality samples of $Li_2Pd_3B$ and $Li_2Pt_3B$ by changing conditions in synthesizing, e.g. heating power and time, sample setting position on the hearth or substituting boron with aluminum. All samples showed a single phase in the XRD analysis. Their quality deteriorations are shown through the measurement of residual resistivity (RR). Electrical resistivity measurements have been done by standard four-probe method under magnetic fields by using the Physical Property Measurement System (PPMS: Quantum Design).

## 3. Experimental results and discussion

Figure 1 shows the temperature dependence of the electrical resistivity in the different quality samples of (a) $Li_2Pd_3B$ and (b) $Li_2Pt_3B$. Superconducting transitions are shown around 7 K for $Li_2Pd_3B$, and 3K for $Li_2Pt_3B$. Sharp transitions and clear zero resistivity is observed in all samples. The aluminum content of these samples are 0% (Pd#4), 1% (Pd#2), 5% (Pd#1), 10% (Pd#3), respectively. Their quality deteriorations are shown through the measurement of residual resistivity (RR). The Pd#1 with smallest RR is expected to include the least impurities and defects. Contrary to expectations, the Pd#4 (no aluminum content) does not show the smallest RR. In synthesis process of these compounds, it is difficult to control the homogeneity of samples, because very short time melting process and low arc power are required to prevent Li loss. If the melting point of $Pd_3B$ or $Pt_3B$ alloy becomes lower, Li loss becomes less in synthesis processes. We confirmed in a preliminary measurement that the melting point of $Pd_3$ $(B_{0.95}Al_{0.05})$ alloy is the lowest among the $Pd_3$ $(B_{1-x}Al_x)$. This is probably a reason why the Pd#1 has the smallest RR. Four samples were also prepared for $Li_2Pt_3B$ and Pt#1 is the highest quality sample among them. The aluminum processes. We confirmed in a preliminary measurement that the melting point of $Pd_3$ $(B_{0.95}Al_{0.05})$ alloy is the lowest among the $Pd_3$ $(B_{1-x}Al_x)$. This is probably a reason why the Pd#1 has the smallest RR. Four samples were also prepared for $Li_2Pt_3B$ and Pt#1 is the highest quality sample among them. The aluminum content of these samples are 0% (Pt#3), 0% (Pt#4), 1% (Pt#2), 5% (Pt#1), respectively. Figure 2(a) shows the temperature dependence of the upper critical field $H_{c2}$ in four samples of $Li_2Pd_3B$. The critical temperatures $T_c$ were determined at the mid point between normal and zero resistivity. The values of $T_c$ are slightly decreased, while the $H_{c2}(0)$ value clearly increased about 1.5 times in low quality sample Pd#4. We discussed this behavior in ref. [25]. It has a similar behavior in ordinary s-wave superconducting state explained by nonlocal generalization of the London equation introduced by Pippard [24]. The coherence length $\xi$ in the presence of scattering were assumed to be related to that of pure material $\xi_0$ by [24]

$$\frac{1}{\xi} = \frac{1}{\xi_0} + \frac{1}{\alpha \ell} \quad (1),$$

where $l$ is the mean free path and $\alpha$ is a numerical constant [25]. From the relation of the orbital limit equation $H_{c2}(0) = \phi_0/2\pi\xi^2$ ($\phi_0 = h/2e$) is the quantum fluxoid), we can expect $H_{c2}(0)$ to increase when $l$ becomes short. We can estimate the coherence length of $\xi_0$ in an ideally pure material from Fig.3 under the assumption of $(\alpha \ell)^{-1} \propto \rho_{RR}$. The value of $\xi_0$ is obtained from the y-intercept of the $\xi^{-1}$ vs. $\rho_{RR}$ plot in Fig.3. The value of $\xi_0$ in ideal quality sample was estimated as 11.6 nm for $Li_2Pd_3B$ [24].

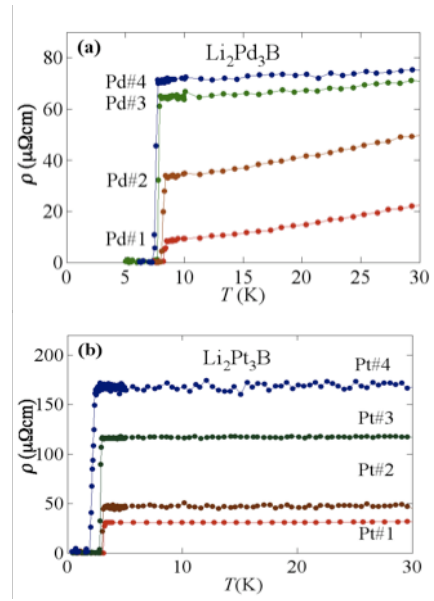

**Figure 1.** $T$ dependence of electrical resistivity (a) in $Li_2Pd_3B$, (b) in $Li_2Pt_3B$. They show sharp transition and clear zero resistivity in four different quality samples of $Li_2Pd_3B$ and $Li_2Pt_3B$. The Pd#1 and Pt#1 with smallest RR are expected to include the least impurities and defects. Pd#4 and Pt#4 are the lowest quality samples.

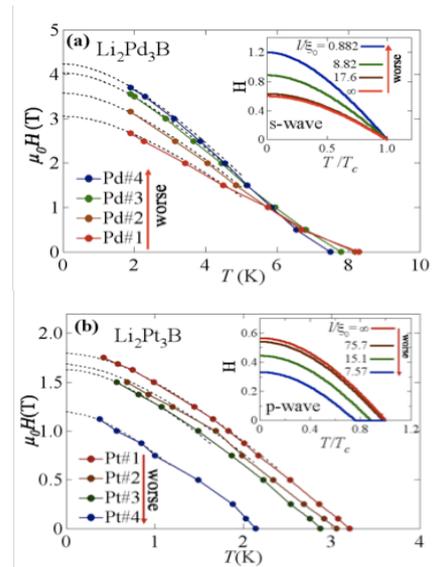

Figure 2. H-T phase diagram, (a) in $Li_2Pd_3B$, (b) in $Li_2Pt_3B$. The inset shows H-T phase diagram of s-wave and $p_z$-wave in theoretical calculations. $Li_2Pd_3B$ exhibits the small $T_c$ suppression attributed by the non-magnetic impurities and defects, while the $H_{c2}(0)$ value clearly increased in the low quality samples. It resembles the theoretical calculations result of s-wave superconducting state. The $T_c$ and $H_{c2}$ suppressions by the non-magnetic impurities and defects are clearly observed in the spin triplet dominant parity mixing superconductor $Li_2Pt_3B$ in contrast to $Li_2Pd_3B$. It also resembles the theoretical calculations result of non s-wave superconducting state.

The mean free path $l$ is obtained from

$$l = 1.00965 \times 10^{-19} (\xi_0 \gamma T_c \rho_{RR})^{-1} \quad (2),$$

where $\xi_0$ (m) is the BCS coherence length, $\gamma$ ($Jm^{-3}K^{-2}$) is the electronic specific-heat coefficient, $T_c$(K) is critical temperature and $\rho_{RR}$ ($\Omega m$) is the residual resistivity [25,26]. In a low quality sample, the mean free path $l$ is shorter and the coherence length $\xi$ becomes short. This H vs. T phase diagram also resembles the theoretical calculation in ordinary s-wave superconducting state as shown in Fig.2 (a) inset.

Figure 1 (b) shows the results in $Li_2Pt_3B$. Sharp transitions and clear zero resistivity is observed to be the same as $Li_2Pd_3B$ shown in Fig.1 (a). The H-T superconducting phase diagram in Fig.2 (b) shows the different aspect to that in $Li_2Pd_3B$. Both of $H_{c2}(0)$ and $T_c$ values are decreased by non-magnetic impurities and defects. It is suggested that the Cooper pair is broken and the superconducting gap is suppressed by non-magnetic impurities and defects. A phase diagram calculated for a $p_z$-wave superconducting state

On the cylindrical Fermi surface is shown in Fig.2 (b) inset. $Li_2Pt_3B$ with cubic crystal structure does not have a simple cylindrical Fermi surface, however the theoretical calculation phase diagram is a reference in the case of a spin triplet line-node superconducting gap. In this case, superconducting energy gap is suppressed by nonmagnetic impurities and both $H_{c2}(0)$ and $T_c$ values are decreased. The H vs. T superconducting phase diagram in the spin triplet dominant parity mixing superconductor $Li_2Pt_3B$ is similar to that in unconventional superconductors. These results strongly indicate that $Li_2Pt_3B$ is an unconventional superconductor in contrast to $Li_2Pd_3B$.

The values of $\xi^{-1}$ and the mean free path $l$ are plotted against residual resistivity (RR) in the four different quality samples of $Li_2Pd_3B$ in Fig.3. The values of coherence length $\xi^{-1}$ are 11.0nm (Pd#1), 9.2nm (Pd#2), 8.2nm (Pd#3), 7.7nm (Pd#4), respectively. The values of $l$ estimated from Eq. (2) are 63.7 nm (Pd#1), 15.9nm (Pd#2), 8.3nm (Pd#3), 7.6 nm (Pd#4), respectively. The $l$ value in the lowest quality sample (Pd#4) is shortened about 1/8 of the best sample (Pd#1). The value of $\ell/\xi_0$ is decreases to about 0.7 in Pd#4, superconductivity is gradually deviating from the clean limit condition in the low quality samples.

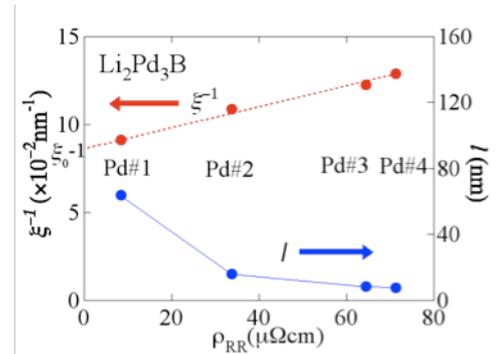

Figure 3. The residual resistivity (RR) dependence of $\xi^{-1}$ and the mean free path $l$ in $Li_2Pd_3B$. $\xi^{-1}$ and the mean free path $l$ plotted against residual resistivity (RR) in the four different quality samples. The value of $\xi_0$ in ideal quality sample is obtained from the y-intercept of the $\xi^{-1}$ vs. $\rho_{RR}$ plot

The coherence length $\xi$ and the mean free path $l$

are plotted against RR in Li$_2$Pt$_3$B in Fig.4. The values of coherence length $\xi$ are 13.5nm (Pt#1), 13.9nm (Pt#2), 14.2nm (Pt#3), 16.6nm (Pt#4), respectively. The values of $l$ are also estimated from Eq. (2), they are 49.1 nm (Pt#1), 31.6 nm (Pt#2), 12.9 nm (Pt#3), 9.1 nm (Pt#4), respectively.

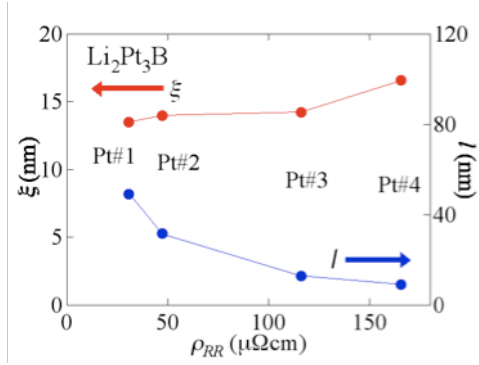

Figure 4. The residual resistivity (RR) dependence of the coherence length $\xi$ and the mean free path in Li$_2$Pt$_3$B. The $\xi$ and $l$ plotted against residual resistivity (RR) in the four different quality samples.

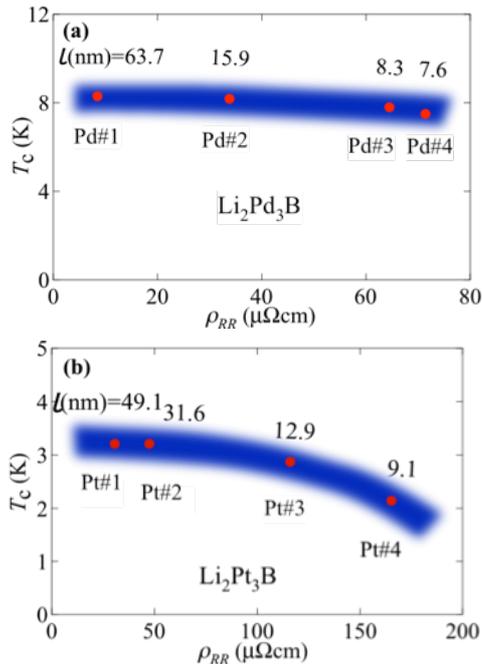

Figure 5. The residual resistivity (RR) dependence of $T_c$, (a) in Li$_2$Pd$_3$B, (b) in Li$_2$Pt$_3$B. $T_c$ values are plotted against the RR. Small $T_c$ suppression was shown in Li$_2$Pd$_3$B, while clear suppression was observed in Li$_2$Pt$_3$B. The values of the mean free path $l$ are indicated in figures.

The $T_c$ values are plotted against the RR in Fig.5, (a) in Li$_2$Pd$_3$B and (b) in Li$_2$Pt$_3$B. The values of mean free path $l$ are also indicated in figures. Although $T_c$ slightly decreases in the low quality samples, we can recognize that Li$_2$Pd$_3$B is not weak against the nonmagnetic impurity. In the case of a conventional s-wave superconductor with anisotropic superconducting energy gap, $T_c$ is slightly affected by even non-magnetic impurities. The slightly suppression in $T_c$ values of Li$_2$Pd$_3$B is probably caused by the complex superconducting gap of Li$_2$Pd$_3$B. For another reason, it may be caused by broken the small amount of spin triplet component of the parity mixing superconductor Li$_2$Pd$_3$B. In any case Li$_2$Pd$_3$B shows a conventional s-wave-like behavior. In Fig.5 (b), $T_c$ suppressions are clearly shown in Li$_2$Pt$_3$B in contrast to Li$_2$Pd$_3$B in Fig. 5 (a).

On comparison of the *H-T* phase diagram in Li$_2$Pd$_3$B and Li$_2$Pt$_3$B, it is proved that non-magnetic impurities and defects cause the different effect on the parity mixing superconductors with different mixing rate. The Cooper pair breaking is clearly caused in the spin triplet dominant (occupying about 2/3 of all) superconductor of Li$_2$Pt$_3$B. The values of $\ell/\xi_0$ decrease to less than 1 in the lowest quality sample. The value is generally too small for an ordinary spin triplet superconductor being sustained. The result resembles the theoretical discussion of impurity effect by Samokhin and mineev [21]. The two-component structure of the order parameter allows for two distinct pairing channels, a dominant and a subdominant gap. In the dirty system only the conventional pairing component would eventually survive, while all alternative pairing channels are suppressed. Even in the spin triplet dominant superconductor of Li$_2$Pt$_3$B, s-wave-like component are expected in 1/3 of all [17]. It may be the reason. It is known that the resistivity is not selective/informative in probing the type or character of the impurities. For the next step, microscopic measurements are needed for detail information to discuss the parity mixing superconducting state on the actual Fermi surfaces. We need detail information of changing the parity mixing ratio, the Fermiology and the scattering mechanism.

## 4. Conclusion

In our results, the *s*-wave dominant parity mixing superconductor Li$_2$Pd$_3$B exhibited the small $T_c$ suppression attributed by the non-magnetic impurities and defects, while the $H_{c2}(0)$ value clearly increased about 1.5 times larger in the low quality samples. These behaviors are similarly observed in ordinary *s*-wave superconductors. The $T_c$ and $H_{c2}$ suppressions by the non-magnetic impurities and defects were clearly observed in the spin triplet dominant parity mixing superconductor Li$_2$Pt$_3$B in contrast to Li$_2$Pd$_3$B. On comparison of the *H-T* phase diagram in Li$_2$Pd$_3$B and Li$_2$Pt$_3$B, it is proved that non-magnetic impurities and defects cause the different effect on the parity mixing superconductors with different mixing rate.

## Acknowledgements

This work was supported by the "Topological Quantum Phenomena" (No. 22103004) Grant-in Aid for Scientific Research on Innovative Areas under the Ministry of Education, Culture, Sports and Science.

## References

[1] G. Dresselhaus, Phys. Rev. Lett. 580 (1955) 100.
[2] E. Rashba, Tverd Fiz (Leiningrad) Tela, Sov. Phys. Solid State 1 (1959) 368.
[3] Gorkvand L P, Rashba E I, Phys. Rev. Lett. 87 (2001) 037004.
[4] P. W. Anderson, Phys. Rev. B 30 (1984) 4000.
[5] V. P. Mineev, Int. J. Mod. Phys. B 18 (2004) 2963.
[6] E. Bauer et al. Phys. Rev. Lett. 92 (2004) 027003.
[7] T. Akazawa, H. Hidaka, T. Fujiwara, T. C. Kobayashi, E. Yamamoto,Y. Haga, R. Settai, and Y. O. nuki, J. Phys. Soc. Jpn. 73 (2004) 3129.
[8] N. Kimura, K. Ito, K. Saitoh, Y. Umeda, H. Aoki, and T. Terashima, Phys. Rev. Lett. 95 (2005) 247004.
[9] N. Kimura, Y. Muro, and H. Aoki, J. Phys. Soc. Jpn. 76 (2007) 051010.
[10] I. Sugitani, Y. Okuda, H. Shishido, T. Yamada, A.Thamizhavel, E.Yamamoto, T. D. Matsuda, Y. Haga, T. Takeuchi, R. Settai, and Y.O. nuki, J.Phys. Soc. Jpn. 75 (2006) 043703.
[11] R. Settai, T. Takeuchi, Y. O. nuki, J. Phys. Soc. Jpn. 76 (2007) 051003.
[12] R. Settai: private communication; See also, A. Thamizhavel, H. Shishido, Y. Okuda, H. Harima, T. D. Matsuda, Y. Haga, R. Settai, and Y. O. nuki, Phys. Soc. Jpn. 76 (2007) 051010, (2006) 044711.
[13] U. Eibenstein ,W J.Jung, Solid State Chem.133 (1997) 21.
[14] P. Badica, T. Kondo, K. Togano, J. Phys. Soc. Jpn. 74 (2005) 1014-1019.
[15] M. Nishiyama, Y. Inada, G.-q. Zheng, Phys. Rev. B 71 (2005) 220505.
[16] M. Nishiyama, Y. Inada, G.-q. Zheng, Phys. Rev. Lett. 98 (2007) 047002.
[17] H. Q. Yuan, D. F. Agterberg, N. Hayashi, P. Badica, D. Vandervelde, K. Togano, M. Sigrist, and M. B. Salamon, Phys. Rev. Lett. 97 (2006) 017006.
[18] H. Takeya, M. ElMassalami, S. Kasahara and K. Hirata, Phys. Rev. B 76 (2007) 104506.
[19] R. Khasanov, I. L. Landau, C. Baines, F. La Mattina, A. Maisuradze, K. Togano, H. Keller, Phys. Rev. B 73 (2006) 214528.
[20] S. Harada, J. J. Zhou, Y. G. Yao, Y. Inada, G-Q. Zheng, Phys. Rev. B 86 (2012) 220502.
[21] V. P. Mineev1 and K. V. Samokhin, PHYSICAL REVIEW B 75 (2007) 184529.
[22] P. A. Frigeri, D. F. Agterberg, I. Milat and M. Sigrist: cond- mat/0505108.
[23] K. Togano, P. Badica, Y. Nakamori, S. Orimo, H. Takeya, K. Hirata: Phys. Rev. Lett. 93 (2004) 247004.
[24] Pippard A B, Proceedings the Royal Society A. 216 (1953) 547-568.
[25] G Bao, Y Inada, S Harada and G-q Zheng, J. Phys: Conf.Ser.391 (2012) 012084.
[26] T. P. Orlando *et.al.*, Phys.Rev.B**19 (**1979) 4545
[27] P. Badica, T.Kondo, et al., Appl.phys.Lett. 93 (2004) 247004.